\documentstyle{article}
\def\ph{{\it\Phi}}

\def\zz{\varphi}

\def\ea{{ et\thinspace al.}\ }                        
\def\eg{{\it e.g.}\ }                                    
\def\ie{{\it i.e.}}                                   

\def\gsim{\lower.4ex\hbox{$\;\buildrel >\over{\scriptstyle\sim}\;$}}
\def\lsim{\lower.4ex\hbox{$\;\buildrel <\over{\scriptstyle\sim}\;$}}

\def\apj{{\it ApJ}\ }

\def\apjl{{\it ApJ Letters}\ }

\def\phr{{\it Phys. Rev.}\ }
\def\phrl{{\it Phys. Rev. Lett.}\ }
\def\phl{{\it Phys. Lett.}\ }

\def\SG{Stefan Gottl\"ober}
\def\sg{Gottl\"ober}
\def\JM{Jan P. M\"ucket}
\def\jm{M\"ucket}

\def\AAS{Alexei A. Starobinsky}

\begin{document}

\title{  Confrontation of a Double Inflationary Cosmological Model with
Observations}

\author{\SG, \JM, \\
Astrophysikalisches Institut Potsdam,\\ An der
Sternwarte 16,\\ D-14482 Potsdam, Germany \\
and \\
\AAS, \\
Landau Institute for Theoretical Physics,\\
Kosygina St. 2, \\
Moscow, 117334, Russia}

\maketitle

\begin{abstract}
CDM models with non-scale-free step-like spectra of adiabatic
perturbations produced in a realistic double inflationary model are
compared with recent observational data. The model contains two
additional free parameters relatively to the standard CDM model with the
flat ($n=1$)
initial spectrum. Results of the COBE experiment are used for the
determination of a free overall spectrum normalization. Then predictions
for the galaxy biasing parameter, the variance for "counts in cells", the
galaxy angular correlation function, bulk flow peculiar velocities
and the Mach number test are obtained. Also considered are conditions
for galaxy and quasar formation. Observational data strongly
restricts allowed values for the two remaining model parameters. However,
a non-empty region for them satisfying all considered tests is found.
\end{abstract}

\section{Introduction} \label{intr}

   Inflationary cosmological models (Starobinsky 1980, Guth 1981, Linde 1982,
1983) imply a density parameter $\Omega_{tot} \approx 1$
($\mid\Omega_{tot} - 1\mid < 10^{-4}$) within the observable part of
the Universe. Combined with the value $\Omega_{bar}\approx 0.017h^{-2}$
following from the theory of primordial nucleosynthesis (see, \eg Walker
\ea 1991) this requires the most of matter in the Universe to be
nonbaryonic ($h=H_0/100$ km/s/Mpc). Second, the  simplest inflationary
models (with one slowly
rolling effective scalar field) predict
density fluctuations with a spectrum of approximately Zeldovich-Harrison
type (i.e., $({\delta \rho \over \rho})_k^2 \propto k^n$ with
$n\approx 1$). This result was first consistently derived
by Hawking 1982, Starobinsky 1982, and Guth
\& Pi 1982. The cold dark matter (CDM) model with this initial spectrum
of perturbations and with biasing $b_g \simeq 1.5 - 2.5$ have most
successfully explained the observed hierarchy of cosmic
structures up to scales of approximately 10 $h^{-1}$ Mpc (Davies \ea
1985).  However, observations of such structures as the Great Attractor
(Lynden-Bell \ea 1988, and Dressler 1991) and the Great Wall
(de Lapparent \ea 1986), the large-scale clustering in the redshift
 survey of IRAS galaxies
(Efstathiou \ea 1990a, Saunders \ea 1991), and the galaxy angular
correlation function and "counts in cells" for
the deep  APM galaxy survey (Maddox \ea 1990, see also Loveday \ea 1992)
imply that there is more power in the perturbation spectrum at scales
larger than approximately $ 10 h^{-1}$ than expected in the standard
model (biased CDM plus $n = 1$ initial spectrum of adiabatic
perturbations). On the other hand, it is remarkable that the standard model
is so close to these observational data: only a modest increase in
amplitude of perturbations at large scales is required to fit them - no
more than 2 - 3 times, and the latest results have the tendency to diminish
this number (see e.g. Loveday \ea 1992 ). Finally, recent COBE
measurements (Smoot \ea 1992) also imply the ratio of amplitude at
scales $(10^3 - 10^4) h^{-1} $ Mpc to that at $(1 - 10) h^{-1}$ Mpc equal to
$(1.1 \pm 0.2) b_g$ ($1\sigma $ error bars). The earlier published positive
RELICT - 1 result
for $\Delta T/T$ at large angles (Strukov \ea 1992a,b) is even larger but
it was obtained using only one wavelength, so it may be at least partially
non-primordial.

The spectrum of perturbations observed at the present time is a product of
an initial (primordial) spectrum and some transfer function $T(k)$. The
latter results from a transition from the radiation-dominated era to the
matter dominated one at redshifts $z \sim 10^4$ and depends on the
structure of dark matter. Thus, any deviation from the standard model
may be explained either by changing the initial flat spectrum, or by
having a different $T(k)$ due to a more complicated matter content.
The latter possibility arises, for example, in models with a mixture
of hot and cold dark matter (Shafi \& Stecker 1984, Holtzmann 1989,
van Dalen \& Schaefer 1992, and others) or with a cosmological constant
and dark matter
(Peebles 1984, Kofman \& Starobinsky 1985, Efstathiou \ea 1990b, and Gorski
\ea 1992). For a recent reanalysis of the former model see Pogosyan \&
Starobinsky 1993; the same for the latter model - see Bahcall \ea 1993.

Here, we consider CDM models with non-scale-free primordial perturbation
spectra.
The scale invariance of the perturbation
spectrum can be broken by different physical mechanisms during
an inflationary stage (see, e.g., Kofman \ea 1985, Salopek \ea, 1989).
Scale-free but not scale-invariant spectra with $n < 1$ seem not to be able
both to provide enough power at $L \sim (25-50)h^{-1} $ Mpc
and to fit the COBE
data (Polarski \& Starobinsky 1992, Liddle \ea 1992, Adams \ea 1993;
see, however, Cen \ea 1992). A
non-scale-free initial spectrum with an effective step ( compared to
the $n=1$ spectrum) somewhere between
$1 h^{-1} $ Mpc and $10 h^{-1} $ Mpc works much better.

To obtain such a spectrum, it is necessary to abandon at least one of the
main assumptions of the simplest version of the inflationary scenario, i.e.
to assume either that the slow-rolling condition is temporarily violated at
some moment of time (and then a step of an universal form in the spectrum
arises, see Starobinsky 1992), or that there is more than one effective
scalar field driving inflation. The latter possibility leads to double
inflation , i.e., to
a cosmological model with two
subsequent inflationary stages (Kofman \ea 1985, Silk \ea
1987). These inflationary stages may be driven by the $R^2$ term and a
scalar field (\sg \ea 1991), where $R$ is the Ricci scalar or by two
noninteracting scalar fields (Polarski \& Starobinsky 1992). In this case, a
step in the spectrum arises as a result of a rapid decrease of the
Hubble parameter $H=\dot a/a$ in a period between two inflationary
stages with slowly varying $H$ and $\dot H$. Further, we consider the
first of the above mentioned double inflationary models, results for the
second one are expected to be similar.

\section{Perturbation spectrum and normalization} \label{pert}

The Lagrangian density of the
gravitational field including the $R^2$ term and a massive scalar field
reads

\begin{equation}
L = {1\over 16\pi G}(-R + {R^2\over 6M^2}) + {1\over
2}(\varphi_{,\mu }\varphi^{,\mu } - m^2\varphi^2).
\label{lag}\end{equation}

\medskip\noindent
The $R^2$ term is coupled via $M^2 \ll G^{-1}$ to General
Relativity, $\varphi$ is the scalar
field and $m^2 \ll G^{-1}$ is its mass ($c = \hbar = 1$). The model
contains two more parameters than the standard model. They define the
location and the relative height of an effective step in the
perturbation spectrum. The
height of the step depends on the ratio $M/m$ and the energy
density of the scalar field at the onset of the second
inflation. On the other hand, this energy density is responsible
for the length of the second inflationary stage, \ie, for the
physical scale at which a break in the spectrum appears.  With
$\zz = \zz_0 \approx 3G^{-1/2}$ at the end of the first inflationary
stage, the break occurs just at the right place,
and the height of the step is $\Delta\approx M/6.5 m$. Note that due
to the exponential dependence of the break location on the energy
density, its shifting from
10 Mpc to 100 Mpc does not practically
influence the height of the step.

We consider the rms value of the Fourier transform of an
gravitational perturbation $\Phi(k)$. Here the conventions
 $\Phi(\vec k)=(2 \pi)^{-3/2}\int \, \Phi(\vec r)e^{-i\vec
k\vec r}\, d^3r$, $\langle \Phi(\vec k) \rangle =0$,
$\langle \Phi(\vec k)\Phi^{\ast}(\vec k')\rangle =
\Phi^2(k)\delta^3(\vec k -
\vec k'),\,\, k=|\vec k|$ for initially Gaussian perturbations are
adopted. Provided the limit $k \to 0$ of $\ph (k)$ coincides with a
 flat spectrum $\tilde\ph (k)$. Then we can construct the two quantities
$\ph^{1} = \lim_{k \to k_{br}-0} \tilde\ph (k)$ and
$\ph^{2} = \lim_{k \to k_{br}+0} \ph (k)$. The height of the step
$\Delta$ is given by  $\Delta =\ph^{(1)}/\ph^{(2)}$. In this paper we consider
spectra with steps $\Delta \approx M/6.5m =
2,\; 3,\; 4,\; 5$.
  The step is located
at $L \simeq 2\pi k^{-1}_{br}$,
where $k^{-1}_{br}$ denotes the wave number where the perturbation spectrum
reaches the lower plateau at large $k$, the values $k^{-1}_{br} =$
(1,3,7,10,20,30) Mpc$^{-1}$ were considered. Here and further
throughout the paper, unless a
scaling by $h$ is given, we assume $h = 0.5$.
The length of a
transition period between flat parts of the spectra is of the order of one
magnitude of wave numbers.

\unitlength1cm
\begin{figure}
\begin{picture}(12,7)
\includegraphics{apj1.ps}
\end{picture}
\caption[]{Density perturbation spectra $P_i$ (i =1,2,3,4) for $M=13m,
M=20m, M=26m, M=32.3 m$ respectively, normalized to $\sigma_T(10^{\circ})$
by the use of the COBE data.
The dashed line shows the standard CDM model with the flat initial
spectrum normalized in the same way.}
\label {fig1}
\end{figure}

Results of numerical calculations of the power spectra of density
perturbations at the present epoch

\begin{equation}
P(k)\equiv P(k,z=0)=\left({\delta \rho \over \rho}\right)_{\vec k}^2
={1\over 36}(kR_H)^4\Phi^2(k)T^2(k)
\end{equation}

\noindent
are plotted in Fig.\ref{fig1}.
Here $T(k)$ is the transfer function for the standard CDM model
(a review of best fits to it see in Liddlle \& Lyth 1993),
for equations for perturbations see Gottl\"ober \ea 1991,
the overall normalization is explained below. We adopt
$\Omega_{tot}=1$ in accordance with inflation, thus $a(t)\propto
t^{2/3}$ now. $R_H=2H_0^{-1}\approx 6000h^{-1}$ Mpc is the present day
Hubble radius sometimes
non-rigorously called the cosmological horizon. For $\Delta \le 3$,
there is practically no oscillations in the spectrum. Then a very good
analytical approximation for it may be obtained by assuming that the
slow-roll condition is valid for both inflaton fields $R$ and $\phi$
in the region of the break (see, e.g., Gottl\"ober \ea 1991).

Let us present this solution in a different form. The background equations
for the model (1) in the slow-roll approximation

\begin{eqnarray}
3H\dot \phi + m^2\phi =0, \\
{6H^2\dot H\over M^2}={4\pi Gm^2\phi^2\over 3}-H^2
\end{eqnarray}

\noindent
have an integral

\begin{equation}
s\equiv -\ln {a\over a_f}= 2\pi G\phi^2+{3H^2\over M^2},
\end{equation}

\noindent
where $a_f$ is the value of the scale factor at the end of inflation (Kofman
\ea 1985, Starobinsky 1985a). Then, introducing $f=H^2/s$, we arrive to the
equation:

\begin{eqnarray}
{df\over d\ln s}={1\over f}(f-f_1)(f_2-f),   \\
f_1={2\over 3}m^2,\; f_2={1\over 3}M^2,\; f(s)<f_2. \nonumber
\end{eqnarray}

\noindent
The scalar field follows from the relation:

\begin{equation}
2\pi G\phi^2 = s\left( 1-{f(s)\over f_2}\right) .
\end{equation}

\noindent
Let, e.g., $M>m\sqrt 2$. Then the solution of Eq.(6) is

\begin{equation}
{(f_2-f)^{{f_2\over f_2-f_1}}\over |f-f_1|^{{f_1\over f_2-f_1}}}
={s_1\over s}(f_2-f_1).
\end{equation}

\noindent
$f$ may be both larger and smaller than $f_1$. $f={f_1+f_2\over 2}$ for
$s=2s_1$. For values of the quantities $M/m$ and $\phi$ used in our
calculations, $f_1\ll f_2$ and $s_1\gg 1$. Using the general
expression for adiabatic perturbations generated during a slow-roll
multiple inflation (Starobinsky 1985a) we get the answer:

\begin{equation}
k^3\Phi^2(k)={9\over 100}\cdot 16\pi Gs^2f(1-{f\over 2f_2}),
\end{equation}

\noindent
where $s$ and $f(s)$ are taken at the moment of horizon crossing at the
inflationary stage $k=a(t)H(t)$; so $s=s(k)\approx \ln {k_f\over k}$ in
this expression, $k_f=a_fH_f$.

It is useful to note here that the background dynamics of an
inflationary model with two massive scalar fields (see Polarsky \&
Starobinsky 1992) in the slow-rolling regime may be reduced to the same
equation (6) with $f=H^2/s$ and $f_{1,2}={2\over 3}m_{1,2}^2$
($m_1<m_2$). Thus, the replacement rule is $M^2\to 2m_2^2$. The integral
(7) is then changes to $s=2\pi G(\phi_1^2+\phi_2^2)$. $f_1<f<f_2$ in
this case, so inflationary dynamics of the model with two massive
scalar fields is isomorphous to only a finite part of possible
inflationary regimes of the model (1). The expression for the power
spectrum of adiabatic perturbations is slightly different from (9):

\begin{equation}
k^3\Phi^2(k)={9\over 100}\cdot 16\pi Gs^2f.
\end{equation}

Now let us turn to an overall normalization of the power spectra. It is
chosen to fit the COBE data (Smoot \ea 1992) on large-angle
$\Delta T/T$ anisotropy ($2\le l <30$, where $l$ is the multipole
number). If $a_l$ is a rms multipole value summed over all $m$ and
averaged over the sky:

\begin{equation}
a_l^2={1\over 4\pi}\sum_{m=-l}^l\langle \left({\Delta T\over T}\right)_
{lm}^2\rangle
={2l+1\over 4\pi}\langle \left({\Delta T\over T}\right)_{lm}^2\rangle\, ,
\end{equation}

\noindent
then the expected variance of the COBE data $\sigma_T^2(\theta_{FWHM})$
may be expressed as

\begin{equation}
\sigma_T^2(\theta_{FWHM})=\sum_{l\ge 2}a_l^2\exp \left( -l(l+1)
\theta_s^2\right),
\end{equation}

\noindent
where $\theta_s=\theta_{FWHM}/2\sqrt {\ln 4}$ is the Gaussian angle
characterizing smearing due to a finite antenna beam size, as well as
due to an additional Gaussian smearing of raw data. We take
$\sigma_T(10^{\circ})=(30\pm 7.5)\mu K/2.735 K$ (Smoot \ea 1992), so
$\theta_s=4.25^{\circ}\approx 1/13.5$.

We have numerically calculated the rms multipole values $a_l$
for the perturbation spectra plotted in Fig.\ref{fig1} (\sg \&{\jm}  1993).
In addition, it is possible to obtain an analytical expression for
$a_l$. Indeed, if $f_1\ll f\ll f_2$, then Eq.(8) simplifies to
$f\approx f_2(s-s_1)/s_1$ for $0<s-s_1\ll s_1$. Thus, $s_1\approx
\ln {k_f\over k_{br}}$. Then it follows from Eq.(9) that

\begin{equation}
k^3\Phi^2(k)\approx {9\over 100}\cdot 16\pi Gf_2s_1(s-s_1)={9\over
100}A^2\ln {k_{br}\over k}
\end{equation}

\noindent
for $k_{br}e^{-s_1}\ll k\ll k_{br}$ ( here we introduce the quantity A
used in Starobinsky 1983). Using the standard formula for the
Sachs-Wolfe effect which is the only significant one for $2\le l<30$
and the case of adiabatic perturbations,
it is straightforward to derive that

\begin{eqnarray}
a_l^2 = {A^2(2l+1)\over 400\pi}\int_0^{\infty} J_{l+1/2}^2(kR_H){\ln
(k_{br}/k)\over k^2R_H}dk=  \nonumber  \\
{A^2 \over 400\pi^2}{2l+1\over l(l+1)} \left(\ln{(2 l_0)} -1 - \Psi (l) -
{l + {1\over 2}\over l(l+1)}\right),
\end{eqnarray}

\noindent
where $l_0 = k_{br} R_{H}\gg 1$ is the multipole number
corresponding to the step location ($l\ll l_0$) and $\Psi$ is the
logarithmic derivative of the
$\Gamma$-function, the difference between $R_H$ and the radius of the
last scattering surface may be neglected here. Substituting $a_l^2$ into
Eq.(12) and choosing $A$ to get the correct value for
$\sigma_T(10^{\circ})$, we obtain properly normalized power spectra
$P(k)$ which are plotted in Fig.\ref{fig1}.

It should be noted that, for the model considered, the contribution of
primordial gravitational waves (GW) to large-angle $\Delta T/T$
fluctuations is small as compared to the contribution from adiabatic
perturbations (AP) and may be neglected. Really, using formulas derived
in Starobinsky 1985a,b, one gets

\begin{equation}
{a_{l(GW)}\over a_{l(AP)}}\approx 2.5\sqrt{3\over 4s_1\ln {k_{br}\over
k_H}}\approx 0.1
\end{equation}

\noindent
for $1\ll l<30$, the value of this ratio for $l=2$ being $\approx 6\%$
more, $k_H=2\pi R_H^{-1}$. Thus, the account of the GW contribution
increases the total rms value of $a_{l(tot)}=\sqrt
{a_{l(AP)}^2+a_{l(GW)}^2}$ by less than $1\%$.

\section{Comparison with observational data}\label{com}
\subsection{ Biasing parameter}

By help of the density perturbation spectrum $P(k)$ we are able
to compute the variance of the mass fluctuation $\sigma_M^2(R)\equiv
\left({\delta M \over M}(R)\right)^2 $ in a sphere of a given radius $R$
(see e.g. Peebles 1980):

\begin{equation}
\sigma_M^2 (R) = {1\over 2 \pi^2} \int^{\infty}_0 k^2 P(k) W(kR) dk,
\label{sigma1}\end{equation}

\noindent
where the window function $W(kR)$ is expressed as

\begin{equation}
W(kR) = {9\over (kR)^6}(\sin kR - kR \cos kR)^2
\label{window}\end{equation}

\noindent
The quantity $\sigma_M^2$
gives the variance of total matter density
perturbations. It becomes clear now that the distribution of galaxies is
biased with respect to the distribution of total matter, so some biasing
parameter $b_g$ (assumed to be scale-independent for simplicity) has to be
introduced:

\begin{equation}
\xi_g(r)=b_g^2\xi (r),\, \sigma_{Mg}^2(R) = b_g^2 \sigma_M^2(R),
\label{sigma2}\end{equation}

\noindent
where $\xi(r)$ is the matter density  correlation function.
Usually it is adopted that $\sigma_{Mg}^2(8 h^{-1}$ Mpc) = 1, so
$b_g=\sigma_M^{-1} (8 h^{-1}$ Mpc). Comparison of N-body simulations with
the observed $\xi_g(r)$ and with the mean pairwise velocity of galaxies
at $r=1h^{-1}$ Mpc (Davies \ea 1985, for more recent calculations see
e.g. Gelb 1993) shows that $b_g$ probably lies between $2$ and $2.5$
($0.4\le \sigma_M(8h^{-1})\le 0.5$). We adopt rather conservative limits
$1.5<b_g<3$ ($0.33< \sigma_M(8h^{-1})<0.67$). Different bias factors
obtained for the considered spectra are given in Table 1 (one should
not forget about $ 25\% $ error bars at $ 1.5\sigma $ level in all these
quantities due to the error bars in $\sigma_{T}(10^{\circ})$). This test
excludes both the standard model and the $P(4)$ case (apart from
the region $k_{br}^{-1}\approx 3$ Mpc  for the latter). The $P(3)$ case
is marginally
admissible, and a broad range of $k_{br}$ remain allowed for the
$P(1)$ and $P(2)$ cases.

\vskip0.5cm
\begin{table}
\begin{tabular}{|c|l|l|l|} \hline
Spectrum $P_i[k_{br}^{-1}]$ & Bias factor & $\sigma_v(R)$ & $\sigma_v(R)$ \\
depending on step           & $b_g$    & km/s & km/s \\
location at $k_{br}^{-1}$ Mpc & & $R=40h^{-1}$ Mpc & $R=60h^{-1}$ Mpc \\ \hline
P1(03) & 1.60  & 269   & 221   \\
P2(03) & 2.18  & 257   & 213   \\
P3(03) & 2.43  & 259   & 215   \\
P4(03) & 2.19  & 268   & 221   \\ \hline
P1(07) & 1.74  & 255   & 211   \\
P2(07) & 2.77  & 230   & 195   \\
P3(07) & 3.49  & 229   & 195   \\
P4(07) & 3.68  & 242   & 204   \\ \hline
P1(10) & 1.77  & 248   & 205   \\
P2(10) & 2.83  & 215   & 184   \\
P3(10) & 3.59  & 210   & 182   \\
P4(10) & 4.18  & 225   & 193   \\ \hline
P1(20) & 1.76  & 234   & 195   \\
P2(20) & 2.67  & 182   & 159   \\
P3(20) & 3.46  & 168   & 151   \\
P4(20) & 4.48  & 180   & 162   \\ \hline
\end{tabular}\\[1.0ex]

TABLE 1: The bias factors and the velocity variances for the models.
P1, P2, P3 and P4 denote models with $M=13m; 20m; 26m; 32.2m$
corresponding to $\Delta = $2, 3, 4, 5. The step location
is given by $k_{br}^{-1}$.
\end{table}
\subsection { Counts-in-cells}
Further we compare the variance $\sigma_c^2(l)$ for the counts-in-cells
analysis of large-scale
clustering in the redshift survey of IRAS galaxies
(Efstathiou \ea 1990) and in the Stromlo-APM survey (Loveday \ea 1992)
with our predictions obtained for corresponding scales. The variance
$\sigma_c^2(l)$ is related to the two-point galaxy correlation function
according to

\begin{equation}
\sigma_c^2(l) = {1\over V^2}\int \int_{V=l^3} \xi_g(r_{12})dV_1 dV_2 \, .
\label{IRAS1}\end{equation}

\noindent
In terms of the density perturbation spectrum $P(k)$, it can be written
as

\begin{equation}
\sigma_c^2(l) = {1\over 2\pi^2}\int_0^\infty k^2 P(k) W_1(kl) dk,
\label{IRAS2}\end{equation}

\noindent
where
\begin{equation}
W_1(\rho) = 8\int^1_0 dx\int^1_0 dy\int^1_0 dz (1-x)(1-y)(1-z)
{\sin(\rho \sqrt{x^2+y^2+z^2})\over \rho \sqrt{x^2+y^2+z^2}}.
\label{IRASw}\end{equation}

\unitlength1cm
\begin{figure}
\begin{picture}(12,7)
\includegraphics{apj2.ps}
\end{picture}
\caption[]{ The variance $\sigma_c^2(l)$ plotted as a function of cell
size $l$. Points obtained from the counts-in-cells
analysis (Loveday \ea 1992) are plotted with $2\sigma$ error bars.
Lines show predictions of $\sigma_c^2(l)$ computed from our spectra with
the step location at $k_{br}^{-1} = 7$ Mpc. The solid lines correspond to
spectra with different step height as indicated at the lines, the dashed
line corresponds to the standard ("flat") CDM
model. The IRAS data are plotted as quadratic symbols.}
\label {fig2}
\end{figure}

A comparison between $\sigma_c^2(l)$ for galaxies from the Stromlo-APM
survey and IRAS galaxies (see Fig.\ref{fig2}) shows that the biasing parameter
for IRAS galaxies $b_{IRAS}<b_{APM}$ for $l=10h^{-1}$ Mpc  and $b_{IRAS}
\approx b_{APM}$ for larger scales (apart from the point $l=40h^{-1}$ Mpc
where $b_{IRAS}^2\approx 1.5 b_{APM}^2$ that is, however, inside
$2\sigma$ error bars). Predictions for $\sigma_c^2(l)$ for the
considered spectra normalized by the condition
$\sigma_{Mg}(8h^{-1}$ Mpc)=1 are shown in Fig.\ref{fig2}. It is seen that
another normalization condition $\sigma_c(12.5h^{-1}$ Mpc)=1 proposed in
Pogosyan \& Starobinsky 1993 works as well. The Kaiser correction (Kaiser 1987)
was not taken into account. In any case, it is clear that this
correction should be rather small that presents one more argument for a
sufficienly large value of $b_g$.

In order to determine values of the model parameters which give the best
fit to these data, we formally apply the $\chi^2$ test considering the
error bars given in Loveday \ea 1992 as $2\sigma$ ones (in the
logarithmic scale of $\sigma_c(l)$). The results are plotted in Fig.\ref{fig3}.
Taking into account the possibility of a change in the overall
normalization, the estimated number of degrees of freedom is $N=7$,
so we consider fits with $\chi^2<7$ as good and fits with $\chi^2>20$
as unacceptable. The best fit $\chi^2\approx 2.2$ occurs for $\Delta=3$
and $k_{br}^{-1}=3$ Mpc.

\unitlength1cm
\begin{figure}
\begin{picture}(12,7)
\includegraphics{apj3.ps}
\end{picture}
\caption[]{ $\chi^2$ test of variances. The contours in the
$k^{-1}_{br}$-$\Delta$-plane represent the levels
$\chi^2(\Delta, k^{-1}_{br}) =$ 3, 7, 20, respectively.}
\label {fig3}
\end{figure}

\subsection{ Angular correlation function}

Let us consider the predictions of our model for the galaxy angular
correlation function $w(\theta)$. To this aim we put the Fourier
transform of the 3-D correlation function $\xi_g(r)$ into the Limber
equation and find

\begin{equation}
w(\theta)={\int_0^{\infty}\,
kP(k)dk\int_0^{\infty}y^4\phi^2(y)J_0(k\theta y)dy\over 2\pi^2
(\int_0^{\infty}y^2\phi(y)dy)^2}.
\end{equation}

\noindent
We insert our power spectra into Eq.(22) and scale the angular
correlation function to the Lick depth ($y^{\ast}\approx 240 h^{-1}$ Mpc)
with the selection function $\phi = y^{-0.5}\exp
\left(-(y/y^{\ast})^2\right)$. Results of the calculations along with
the recent observational data for $w(\theta)$ for the APM survey
(provided by S. Maddox, see also Maddox \ea 1990) are shown in Fig.\ref{fig4}.
The use of linear approximation in the calculation of $w(\theta)$ is
justified for $\theta >0.5^{\circ}$ only. This is the reason of a
discrepancy between our theoretical curves and the data at small angles
that should not be taken into account. At larger angles up to $\theta
\sim 10^{\circ}$, the agreement with the $\Delta=3$ and $\Delta=4$ cases
is very good (and errors are too large to make any definite conclusion
beyond this angle).

\begin{table}
\begin{tabular}{|c|c|c|c|} \hline
Spectrum $P_i[k_{br}^{-1}]$ & \multicolumn{3}{|c|}{Mach numbers} \\ \cline{2-4}
depending on step   & $R_L = 4 h^{-1}$Mpc& $R_L = 8 h^{-1}$Mpc&
$R_L = 18 h^{-1}$Mpc  \\
location at $k_{br}^{-1}$ Mpc  & & & \\ \hline
P1(03) & 2.76  & 1.48   & 0.76   \\
P2(03) & 3.16 &1.68 &0.86 \\
P3(03) &3.35 & 1.76 & 0.89 \\
P4(03) & 3.21 & 1.69 & 0.85 \\ \hline
P1(07) & 2.80  & 1.51   & 0.79  \\
P2(07) & 3.45  & 1.88  & 0.98  \\
P3(07) & 4.01  & 2.17   & 1.10   \\
P4(07) & 4.18  & 2.19   & 1.07   \\ \hline
P1(10) & 2.78  & 1.51   & 0.79   \\
P2(10) & 3.37  & 1.87   & 1.01   \\
P3(10) & 3.90  & 2.20   & 1.19   \\
P4(10) & 4.51  & 2.44   & 1.22   \\ \hline
P1(20) & 2.68  & 1.46   & 0.77  \\
P2(20) & 2.85  & 1.61  & 0.93   \\
P3(20) & 3.09 & 1.76  & 1.07   \\
P4(20) & 3.87 & 2.25  & 1.35   \\ \hline
\end{tabular} \\[1.0ex]
TABLE 2:  The Mach number test (notation as in Table 1)
\end{table}
\subsection{ Large-scale bulk flows}

Next we investigate our spectra with respect to large-scale bulk
flows (peculiar velocities) of matter. The rms peculiar velocity inside
a sphere of a radius $R$ is given by the expression:

\begin{equation}
\sigma_v^2(R) = {2\over \pi^2R_H^2}\int^\infty_0 \; P(k)
                  W(kR) \exp {(-k^2 R^2_S)}dk,
\label{velocity}\end{equation}
where $W(kR)$ is the window function given in eq. (\ref{window}) and
the length scale $R_S$ characterizes the Gaussian
smoothing of raw observational data.
In the Table 1, the corresponding predictions for our model
are shown. They may be compared with recent data from the POTENT
reconstruction of the 3-D peculiar velocity field: $\sigma_v(40h^{-1}$ Mpc)
$ =(405\pm 60)$ km/s and $\sigma_v(60h^{-1}$Mpc) $=(340\pm 50)$ km/s with
$R_S=12h^{-1}$ Mpc, error bars are $1\sigma$ (Dekel 1992, see also
Bertschinger \ea 1990).

This test presents the most serious problem for the model because
expected values of $\sigma_v$ are lower than in models having
exactly flat spectrum at large scales, e.g. in the CDM+HDM $n=1$
model or in the
CDM model with a step in the initial spectrum produced by one scalar
field (Starobinsky 1992), due to the logarithmic increase of $P(k)$ at
$k\ll k_{br}$, see Eq. (13). On the other hand, for $k_{br}^{-1}\le 7$ Mpc,
they are higher than in tilted models with $n<0.75$ and without
significant contribution of gravitational waves to large-angle $\Delta
T/T$ fluctuations.
It should be mentioned here that bulk flow velocities
obtained from observations contain large uncertainties. Second, it is
fairly possible that actual values of bulk flow velocities at our
location in the Universe are larger than average ones ("cosmic
variance"). So, taking the data at their lower $2\sigma$ limit (this
is $\approx
1.4$ times less than the average values that is permitted by cosmic
variance) and pushing the overall normalization of the model to the
upper $1.5\sigma$ limit of the COBE data, we find that the allowed region
is $k_{br}^{-1}\le 7$ Mpc.

\subsection{ Mach number test}

Now we consider the Mach number test for our model (Ostriker \& Suto
1990). The expression for
the Mach number in terms of the quantities introduced above is given as

\begin{equation}
M^2(R_L) = {\int^{\infty}_0 P(k) \exp(-k^2R_S^2) \exp(-k^2 R_L^2) dk\over
      \int^{\infty}_0 P(k) \exp(-k^2 R_S^2)
      \left(1-(1+{k^2 R_L^2/9})\exp(-k^2 R_L^2)\right) dk},
\label{mach}\end{equation}

\noindent
where $R_S = 5 h^{-1} $ Mpc characterizes the Gaussian smoothing.
Ostriker \& Suto (1990) found the Mach numbers $4.2\pm1.0$, $2.2 \pm 0.5$
and $1.3 \pm 0.4$ at the scales $R_L=4h^{-1}$ Mpc, $8h^{-1}$ Mpc,
$18 h^{-1}$ Mpc, correspondingly. The Mach numbers predicted for our spectra
are given in Table 2. Here, once more, the values of $M^2$ obtained from
observations in our vicinity may be larger than average ones due to the
numerator of Eq. (\ref{mach}), i.e., due to local values of large-scale
bulk velocities being larger than average. This test clearly shows that
$\Delta >2$.

\begin{table}
\begin{tabular}{|c|l|} \hline
Spectrum $P_i[k_{br}^{-1}]$ depending on step   & $\chi^2$\\
location at $k_{br}^{-1}$ Mpc  & \\ \hline
P0     &  18.00 \\ \hline
P1(03) & 13.41   \\
P2(03) & 9.59   \\
P3(03) & 11.08   \\
P4(03) &  10.17   \\ \hline
P1(07) & 14.44   \\
P2(07) & 11.18   \\
P3(07) & 13.69  \\
P4(07) & 28.06   \\ \hline
P1(10) & 15.80   \\
P2(10) & 14.27   \\
P3(10) & 13.69   \\
P4(10) & 20.04   \\ \hline
P1(20) & 19.56   \\
P2(20) & 49.82   \\
P3(20) & 24.17   \\
P4(20) & 17.27   \\ \hline
\end{tabular} \\[1.0ex]
TABLE 3: $\chi^2$-values for the models

\end{table}
\vskip0.5cm

\subsection{ Quasar and galaxy formation}

Finally, we consider the compatibility of our model with the existence
of a sufficient number of large galaxies at the redshift $z=1$ and
host galaxies for quasars at $z=4$. The standard model of quasar formation
assumes that they arise as a result of formation of massive black holes
($M\approx 10^9M_{\odot}$) in nuclei of galaxies with total masses
$(10^{11}-10^{12})M_{\odot}$ (incuding the dark matter component).
We estimate $f(\ge M)$ - the fraction of matter in gravitationally bound
objects with masses beginning from $M$ and higher - using a very simple
though rather crude formula by Press \& Schechter (1974):

\begin{equation}
f(\ge M)= 1-\hbox{erfc}\left({\delta_c\over \sqrt 2 \sigma_M(R,z)}\right),
\label{press}\end{equation}

\noindent
where $M={4\over 3}\pi R^3\rho_0$, $\sigma_M(R,z)=\sigma_M(R)/(1+z)$,
$\sigma_M(R)$ is the rms value of the mass fluctuation in the linear
approximation at the present moment, it is defined in Eqs. (16,17).
The choice of the quantity $\delta_c$ that is equal to the value of
${\delta \rho \over \rho}$ in the linear approximation at the moment when
the considered region collapses
in the course of a fully non-linear evolution is critical for this
approach. For the spherical model, $\delta_c=3(12\pi )^{2/3}/20\approx
1.69$, but attempts to fit results of N-body simulations to the
Press-Schechter formula did not produce a unique answer. It seems that
the best fit
for $\delta_c$ lies between 1.33 and 2.

If we accept the most recent estimates of the mass fraction in hosts of
quasars at $z=4$ (see, e.g. Haehnelt 1993): $f(\ge 10^{11}M_{\odot})\approx
10^{-4}$, when it follows from Eq.(\ref{press}) that the corresponding
rms linear mass fluctuation at the present moment

\begin{equation}
\sigma_M(10^{11}M_{\odot})\approx 1.3\delta_c\approx 2.2\pm 0.5
\label{quas}\end{equation}

\noindent
for $\delta_c$ lying in the range mentioned above. Clearly, this is only
a lower limit on $\sigma_M$ because quasars may form not in all host
galaxies and the presenly observed quasar density at $z=4$ may be less
than the real one. However, due to extreme sensitivity of the expression
(\ref{press}) to a change in $\sigma_M$, the actual value of the latter
cannot be significantly larger than that given in Eq.(\ref{quas}).
A similar estimate was presented in recent paper by Blanchard \ea (1993)
(it is $30\%$ higher for their value $\alpha =1.7$).

Another estimate may be obtained from the fact that large galaxies (or,
at least, a significant part of them) seem to be already existent at
$z=1$. Then, assuming $f(\ge 10^{12}M_{\odot})\ge 0.1$ at $z=1$, we get

\begin{equation}
\sigma_M(10^{12}M_{\odot})\ge 1.2\delta_c\approx 2.0\pm 0.4\, .
\end{equation}

\noindent
Note also that the estimate of Haehnelt (1993): $f(\ge 10^{12}M_{\odot})
\ge 10^{-5}$ at $z=4$ - leads to almost the same result:
$\sigma_M(10^{12}M_{\odot})\ge 1.1\delta_c$.

\unitlength1cm
\begin{figure}
\begin{picture}(12,7)
\includegraphics{apj4.ps}
\end{picture}
\caption[]{ Predicted angular correlation functions compared
with the observational data from the APM survey (the data were kindly
provided by S. Maddox). The solid lines correspond to a spectrum
with  $k_{br}^{-1}
=7$ Mpc, the dashed line shows the prediction for the standard CDM
model }
\label {fig4}
\end{figure}

Now the same quantities may be calculated for our spectra. Analysis of
the results shows that the predicted values of $\sigma_M(10^{11}M_{\odot})$
and $\sigma_M(10^{12}M_{\odot})$ for $\Delta =3$ and 1 Mpc $\le
k_{br}^{-1}\le 20$ Mpc lie just inside the required ranges. For $\Delta =2$
they are probably too high, for $\Delta =4$ - too low (apart from the
points $k_{br}^{-1}=1$ Mpc and 20 Mpc).

\section{Conclusions}\label{con}

We have compared the double inflationary model (1) characterized by two
additional free parameters as compared to the standard model with a
number of observational tests and have found that there is a reasonable
agreement with all considered tests for the following region of the
parameters: 1 Mpc $<k_{br}^{-1}<10$ Mpc (that corresponds to real length
scales $L_{br}\approx (6-60)$ Mpc)
and $2<\Delta <4$. The best fit seems to be given by $k_{br}^{-1}=$ (3-7) Mpc,
$\Delta\approx 3$.
Generally, there is a correlation between a step
size and a break location giving a good fit to the data, i.e., for a
break at larger scales also a
higher step is necessary. However, due to a complicated structure
of the obtained spectra (e.g., oscillations throughout the range of
the break for $\Delta >2 $) this connection is not so straightforward.

The most restricting for the model are large scale bulk flows. To obtain
even a marginal agreement we have to assume that their actual rms values
are at least $\approx 1.4$ times less than the average values of Dekel's
(1992) data (due to the cosmic variance) and that the rms value
of large-angle
CMB temperature fluctuations is not less than the average result of
COBE. These assumptions are certainly crucial tests for the considered
model.

\vskip0.5cm
The work of one of the authors (A.S.) was supported by the Russian
research project "Cosmomicrophysics". A.S. also thanks the Deutsche
Forschungsgemeinschaft (DFG) for the finacial support of his visit to
the Potsdam
Astrophysical Institute. We would like to thank Steven Maddox for sending
us the latest version of his $w(\theta )$ data.
\vfill\eject

\end{document}